\documentclass[conference,letter]{IEEEtran}
\usepackage{amsthm}
\usepackage{amsmath}
\usepackage{amssymb} 
\usepackage{amsmath}
\usepackage{amsfonts}
\usepackage{algorithmic}
\usepackage{graphicx}
\usepackage{cite}

\usepackage{subcaption}
\usepackage{textcomp}
\usepackage{xcolor}
\usepackage{tabularx}
\usepackage{lipsum}
\usepackage{float}
\usepackage{stmaryrd}
\usepackage{flushend}
\usepackage{multirow}

\usepackage{url}
\usepackage{ifdraft}
\usepackage{multirow}
\usepackage{hyperref}

\usepackage{footmisc}

\hypersetup{
    unicode=false,          % non-Latin characters in Acrobat's bookmarks
    pdftoolbar=true,        % show Acrobat's toolbar?
    pdfmenubar=true,        % show Acrobat's menu?
    pdffitwindow=false,     % window fit to page when opened
    pdfstartview={FitH},    % fits the width of the page to the window
    pdftitle={The Title of the Paper},
    pdfauthor={Dr. Ferkan Yilmaz},% author
    pdfsubject={Subject},   % subject of the document
    pdfcreator={Creator},   % creator of the document
    pdfproducer={latex.exe + dvips.exe + ps2pdf.exe}, % producer of the document
    pdfkeywords={XXXX, XXXX, XXXX, XXXX.}, % list of keywords
    pdfnewwindow=true,      % links in new window colorlinks=false, % false: boxed links; true: colored links
    linkcolor=red,          % color of internal links
    citecolor=green,        % color of links to bibliography
    filecolor=magenta,      % color of file links
    urlcolor=cyan           % color of external links
}

\newcommand{\argmax}{\mathop{\mathrm{argmax}}\limits}

\newcommand{\IEEEcompact}{
\fontdimen3\font=0.2ex% inter word stretch
\fontdimen2\font=0.5ex% inter word space
}

\theoremstyle{plain}

\theoremstyle{plain}

\theoremstyle{plain}

\theoremstyle{plain}

\theoremstyle{plain}

\newcommand{\figref}[1]{Fig.\,\protect\ref{#1}}

\newcommand{\secref}[1]{Section\,\protect\ref{#1}}

\newcommand{\ITMsymbol}{$\hfill\square$}

\DeclareMathOperator{\diag}{diag}
\DeclareMathOperator{\CRC}{CRC}

\newcommand{\norm}[1]{\left\lVert#1\right\rVert}

\def\BibTeX{{\rm B\kern-.05em{\sc i\kern-.025em b}\kern-.08em
    T\kern-.1667em\lower.7ex\hbox{E}\kern-.125emX}}

\def\BibTeX{{\rm B\kern-.05em{\sc i\kern-.025em b}\kern-.08em
    T\kern-.1667em\lower.7ex\hbox{E}\kern-.125emX}}
\usepackage{hyphenat}
\IEEEoverridecommandlockouts 
\begin{document}
%-------------------------------------------------------------------------------------------------------------------
\IEEEcompact
%%%%%%%%%%%%%%%%%%%%%%%%%%%%%%%%%%%%%%%%%%%%%%%%%%%%%%%%%%%%%%%%%%%%%%%%%%%%%%%%%%%%%%%%%%%%%%%%%%%%%%%%%%%%%%%%%%%%%%%%%%%%
%% ___________.__  __  .__          
%% \__    ___/|__|/  |_|  |   ____  
%%   |    |   |  \   __\  | _/ __ \ 
%%   |    |   |  ||  | |  |_\  ___/ 
%%   |____|   |__||__| |____/\___  >
%%                               \/ 
\title{Toward 6G Downlink NOMA: CRC-Aided GRAND for Noise-Resilient NOMA Decoding in Beyond-5G Networks
\thanks{This work is supported by Istanbul Technical University BAPS\.{I}S MAB-2023-44565 project. The publication is funded by T\"{U}B\.{I}TAK-B\.{I}LGEM. \vspace{-2mm}}
}
\author{
\IEEEauthorblockN{
     	${}^{(a)}$Emirhan Zor,
        ${}^{(b)}$Bora Bozkurt, and 
        ${}^{(c)}$Ferkan Yilmaz}\\[-3mm]
  	\IEEEauthorblockA{
        ${}^{(a,b,c)}${Istanbul Technical University, Electronics and Communication Engineering, Istanbul, T\"{u}rkiye}\\
       ${}^{(b)}${Communications and Signal Processing Research (H\.{I}SAR) Laboratory, T\"{U}B\.{I}TAK B\.{I}LGEM, 41470, Kocaeli, T\"{u}rkiye}\\        
            \{\!
                ${}^{(a)}$\texttt{zor18},
                ${}^{(b)}$\texttt{bozkurtb19},
                ${}^{(c)}$\texttt{yilmazf}
            \!\}\texttt{@itu.edu.tr}
  	}
%------------------------------------------------------------
\vspace{-10mm}
%------------------------------------------------------------
}
\IEEEoverridecommandlockouts
% \IEEEpubid{\makebox[\columnwidth]{000-0-0000-0000-0/00/\$00.00 ©2025 IEEE \hfill}
% \hspace{\columnsep}\makebox[\columnwidth]{ }}
\maketitle
% \IEEEpubidadjcol
%\IEEEcompact

%%%%%%%%%%%%%%%%%%%%%%%%%%%%%%%%%%%%%%%%%%%%%%%%%%%%%%%%%%%%%%%%%%%%%%%%%%%%%%%%%%%%%%%%%%%%%%%%%%%%%%%%%%%%%%%%%%%%%%%%%%%%
%%   _____ ___.             __                        __
%%  /  _  \\_ |__   _______/  |_____________    _____/  |_
%% /  /_\  \| __ \ /  ___/\   __\_  __ \__  \ _/ ___\   __\
%%/    |    \ \_\ \\___ \  |  |  |  | \// __ \\  \___|  |
%%\____|__  /___  /____  > |__|  |__|  (____  /\___  >__|
%%        \/    \/     \/                   \/     \/
\begin{abstract}
\IEEEcompact
Non-Orthogonal Multiple Access (NOMA) technology has emerged as a promising technology to enable massive connectivity and enhanced spectral efficiency in next-generation wireless networks. In this study, we propose a novel two-user downlink power-domain NOMA framework that integrates a Cyclic Redundancy Check (CRC)-aided Guessing Random Additive Noise Decoding (GRAND) with successive interference cancellation (SIC). Unlike conventional SIC methods—which are susceptible to error propagation when there is low power disparity between users—the proposed scheme leverages GRAND’s noise-centric strategy to systematically rank and test candidate error patterns until the correct codeword is identified. In this architecture, CRC is utilized not only to detect errors but also to aid the decoding process, effectively eliminating the need for separate Forward Error Correction (FEC) codes and reducing overall system overhead. Furthermore, the strong user enhances its decoding performance by applying SIC that is reinforced by GRAND-based decoding of the weaker user’s signals, thereby minimizing error propagation and increasing throughput. Comprehensive simulation results over both Additive White Gaussian Noise (AWGN) and Rayleigh fading channels, under varying power allocations and user distances, show that the CRC-aided GRAND-NOMA approach significantly improves the Bit Error Rate (BER) performance compared to state-of-the-art NOMA decoding techniques. These findings underscore the potential of integrating universal decoding methods like GRAND into interference-limited multiuser environments for robust future wireless networks.
\end{abstract}

%%  ____  __.                                    .___      
%% |    |/ _|____ ___.__.__  _  _____________  __| _/______
%% |      <_/ __ <   |  |\ \/ \/ /  _ \_  __ \/ __ |/  ___/
%% |    |  \  ___/\___  | \     (  <_> )  | \/ /_/ |\___ \ 
%% |____|__ \___  > ____|  \/\_/ \____/|__|  \____ /____  >
%%         \/   \/\/                              \/    \/ 

%%%%%%%%%%%%%%%%%%%%%%%%%%%%%%%%%%%%%%%%%%%%%%%%%%%%%%%%%%%%%%%%%%%%%%%%%%%%%%%%%%%%%%%%%%%%%%%%%%%%%%%%%%%%%%%%%%%%%%%%%%%%
%% .___        __                    .___             __  .__               
%% |   | _____/  |________  ____   __| _/_ __   _____/  |_|__| ____   ____  
%% |   |/    \   __\_  __ \/  _ \ / __ |  |  \_/ ___\   __\  |/  _ \ /    \ 
%% |   |   |  \  |  |  | \(  <_> ) /_/ |  |  /\  \___|  | |  (  <_> )   |  \
%% |___|___|  /__|  |__|   \____/\____ |____/  \___  >__| |__|\____/|___|  /
%%          \/                        \/           \/                    \/ 
\vspace{-1mm}
\section{Introduction}
\IEEEcompact
The exponential growth in connected devices—driven by applications such as the Internet of Things (IoT), mission-critical services, and Ultra Reliable Low Latency Communications (URLLC)—necessitates the development of highly efficient and scalable wireless communication systems. Conventional orthogonal multiple access (OMA) schemes, including time-, frequency-, and code-division methods, allocate distinct resources to individual users to avoid interference. However, these methods face limitations in supporting the massive connectivity and low latency requirements of future systems. In contrast, Non-Orthogonal Multiple Access (NOMA) enables multiple users to share the same resources non-orthogonally, significantly enhancing spectral efficiency and access capacity \cite{survey2020makki}.

%\cite{yang2024noma_grand_am,yang2024noma_softgrand,yang2023interference_aware,umar2024joint_sgrand_polar}.

%%
%% New Paragraph
%%
Power-Domain NOMA (PD-NOMA) is a well-studied variant that distinguishes users via power levels. Typically, users with stronger channel conditions are assigned lower transmission power, while weaker users are allocated higher power. At the receiver, Successive Interference Cancellation (SIC) is used: the user with higher received power is decoded first, and its contribution is subtracted from the composite signal before decoding the next user. Despite its relative simplicity, SIC suffers from error propagation and performance degradation when the interfering signal has a power not significantly dominant over the user's signal \cite{survey2020makki}. Considering the limitations of SIC-based methods in NOMA systems, the robust error correction capabilities of the Guessing Random Additive Noise Decoding (GRAND) decoder can serve as a promising complementary solution. By leveraging its unique approach, GRAND may effectively mitigate issues like error propagation, particularly in scenarios where the interfering signal is not significantly dominant over the intended signal. GRAND decoders approach the problem from a noise-centric perspective: rather than decoding the codeword directly, the algorithm guesses and removes noise sequences from the received signal until a valid codeword is found. In more detail, the decoder systematically ranks the candidate error vectors in descending order of estimated likelihood and then evaluates each candidate sequentially. Variants such as Ordered Reliability Bits GRAND (ORBGRAND) and Soft-GRAND (SGRAND) extend this idea to soft inputs and allow the exploitation of channel reliabilities for improved performance \cite{duffy2019capacity,duffy2022ordered}.

%%
%% New Paragraph
%%
A growing body of recent work has investigated using the GRAND methodology within NOMA schemes to enhance performance in intereference-limited multi-user environments. One of the earliest works, given in \cite{multiuser2023yang}, introduces GRAND aided macrosymbol (GRAND-AM), which is a joint multiuser detection method based on the GRAND methodology. The later study in \cite{yang2023interference_aware} extends the GRAND-AM approach by modifying the multiuser detection in GRAND-AM such that the effect of interferers are negated, resulting in superior performance. In \cite{yang2024noma_grand_am}, the materials presented in \cite{multiuser2023yang, yang2023interference_aware} are collected in a more comprehensive framework and extended further by incorporating the channel gain estimate into the detection. As opposed to the previously mentioned studies which examine single-input-single-output (SISO) NOMA systems, the paper in \cite{yang2024noma_grand_am} integrates GRAND-AM within an uplink multiple-input-multiple-output (MIMO) NOMA system. Specifically, GRAND-AM is adapted to work with MIMO methods such as Alamouti space time block codes (STBC), and a performance gain of around 2.5 dB is demonstrated against the multiuser Vertical-Bell Laboratories Layered Space-Time
(V-BLAST) receiver. Finally, in addition to previous GRAND-AM studies, in \cite{umar2024joint_sgrand_polar} polar coding was employed to demonstrate that SGRAND outperforms classical CRC-aided Successive Cancellation List (SCL) decoding (with a list size of 32). Under NOMA interference, CRC‑aided SGRAND achieves gains of approximately 0.5 dB in additive white Gaussian noise (AWGN) channel and up to 1.5 dB in Rayleigh fading channels, while reducing complexity in mid‑to‑high SNR regimes.

This work builds upon the aforementioned advances by integrating GRAND and its variants into downlink PD-NOMA systems and evaluating their performance under various scenarios. Our contributions are summarized as follows:
\begin{itemize}
\IEEEcompact
% % Item
% %---------------------------------------------------
 \item[\ITMsymbol] We propose a 2-user downlink PD-NOMA transmission with GRAND decoding, enabling enhanced spectral efficiency and reliability, which are key capabilities for ultra-reliable and scalable connectivity in beyond 5G networks. Furthermore, we analyze the proposed system's error performance theoretically.
 
 % %---------------------------------------------------
% % Item
% %---------------------------------------------------
 \item[\ITMsymbol] We examine the performance of the 2 user NOMA system through simulation studies using both the classical GRAND and 1-line ORBGRAND methods. The simulations are conducted for traditional NOMA (termed Pure NOMA), and the proposed GRAND-NOMA and GRAND-NOMA with assistance scenarios, showing the BER curves of the two users. The results are given for both the AWGN channel and the Rayleigh fading channel, offering extensive insight into the robustness and practical viability of the proposed next generation transmission scheme.
 
 % %---------------------------------------------------
% % Item
% %---------------------------------------------------
\item[\ITMsymbol] We further explore the system’s performance by providing BER results showing the impact of the distance of the user to the transmitter and the NOMA power allocation constant assigned to the user. As in before, we provide detailed comments regarding the results and explain the displayed behavior of the system. 
\end{itemize}

%%
%% New Paragraph
%%
The remainder of this paper is organized as follows. \secref{Sec:SystemModel} introduces the system model of 2-user downlink NOMA scenario and explores the application of GRAND to improve the performance. \secref{Sec:PerformanceAnalysis} presents the theoretical performance analysis of the proposed scheme. In \secref{Sec:PerformanceAnalysisAndSimulationResults}, performance evaluation of the proposed GRAND-NOMA scheme is conducted by presenting comprehensive simulation results. Finally, \secref{Sec:Conclusion} concludes the key findings and their implications.

\section{System Model}\label{Sec:SystemModel}
\IEEEcompact
This section presents a two-user downlink NOMA model where CRC-encoded messages are superimposed, and the strong user employs GRAND within SIC to remove interference.

%%   _________    ___.                         __  .__               
%%  /   _____/__ _\_ |__   ______ ____   _____/  |_|__| ____   ____  
%%  \_____  \|  |  \ __ \ /  ___// __ \_/ ___\   __\  |/  _ \ /    \ 
%%  /        \  |  / \_\ \\___ \\  ___/\  \___|  | |  (  <_> )   |  \
%% /_______  /____/|___  /____  >\___  >\___  >__| |__|\____/|___|  /
%%         \/          \/     \/     \/     \/                    \/ 
\subsection{2 User Downlink NOMA}
\IEEEcompact
Let us consider a downlink SISO transmission scenario employing PD-NOMA, hereafter referred to as NOMA, involving a single transmitter and two users. Let the message bits of user 1 and that of user 2 be denoted by $\mathbf{u}_1\!\in\!\mathbb{F}_2^K$ and $\mathbf{u}_2\!\in\!\mathbb{F}_2^K$, respectively. Unlike conventional systems that rely FEC codes, we uniquely employ CRC not only for error detection but also as the sole error-correcting mechanism, eliminating the need for dedicated FEC layers. As such, these $K$-bit messages are safeguarded by a CRC code. The CRC encoder is modeled as an injective function, that is $\mathrm{CRC}\colon\mathbb{F}_2^K\rightarrow\mathbb{F}_2^N$, mapping each $K$-bit message into an $N$-bit codeword. The codewords for both users are expressed as $\mathbf{c}_1\!=\!\mathrm{CRC}(\mathbf{u}_1)\!\in\!\mathcal{C}$ and $\mathbf{c}_2\!=\!\mathrm{CRC}(\mathbf{u}_2)\!\in\!\mathcal{C}$, with the CRC codebook $\mathcal{C}\!\equiv\!\left\{\mathbf{c}\mid\mathbf{c}=\mathrm{CRC}(\mathbf{b})\allowbreak~\text{with}~\allowbreak\mathbf{b}\in\mathbb{F}_2^K,\allowbreak\mathbf{c}\in\mathbb{F}_2^N\right\}$.
The corresponding code rate is given by $R\!=\!K/N$. Accordingly, $\forall{i}\!\in\!\{1,2\}$, codeword $\mathbf{c}_{i}$ can be expressed as $\mathbf{c}_{i}\!=\![{c}_{i,1},\allowbreak{c}_{i,2},\ldots,\allowbreak{c}_{i,N}]^T$, where superscript $T$ denotes transposition. These codewords are modulated using a $q$-bit constellation $\mathcal{X}\subset\mathbb{C}$ of modulation order $|\mathcal{X}|$, with each symbol carrying $q\!=\!\log_2|\mathcal{X}|$ bits. As such, codewords $\mathbf{c}_1 \in \mathcal{C}$ and $\mathbf{c}_2 \in \mathcal{C}$ are modulated into $M\!=\!N/q$ symbol sequences $\mathbf{s}_1\!\in\!\mathcal{X}^M$ and $\mathbf{s}_2\!\in\!\mathcal{X}^M$, respectively, with equiprobable symbols drawn from $\mathcal{X}$. Further, the constellation is normalized such that the average power satisfies $\mathbb{E}[\mathbf{s}_i\mathbf{s}_i^*]\!=\!M$ for $i\!\in\!\{1,2\}$. To the best of our knowledge, by giving up the FEC, the proposed CRC-aided NOMA framework, which is the focus of this work, improves the error rate performance and spectral efficiency of NOMA systems not only by reducing signaling overhead but also maximizing throughput while maintaining reliability through CRC-based correction during SIC as explained in \secref{subsec:application_of_GRAND}.

%%
%% New Paragraph 
%%
In a downlink NOMA setting, the transmitted total signal $\mathbf{s}_{\Sigma}\!\in\!\mathbb{C}^M$ is formed by superimposing the symbol sequences of user 1 and user 2, denoted by $\mathbf{s}_1\!\in\!\mathbb{C}^M$ and $\mathbf{s}_2\!\in\!\mathbb{C}^M$, respectively, where $M\!=\!N/q$. Thence, the transmitted total signal $\mathbf{s}_{\Sigma}$ can be expressed as
\begin{equation}
    \mathbf{s}_{\Sigma}=\sqrt{\alpha_1{P}}\,\mathbf{s}_1+\sqrt{\alpha_2{P}}\,\mathbf{s}_2,
\end{equation}
where $P\in\mathbb{R}_{+}$ denotes the power of the transmitted total signal, and $\alpha_1\!\in\!(0,1)$ and $\alpha_2\!\in\!(0,1)$ are the power allocation coefficients satisfying $\alpha_1+\alpha_2\!=\!1$. We consider transmission over generalized fading channels. Consequently, the received signals $\mathbf{r}_1\!\in\!\mathbb{C}^M$ at user 1 and $\mathbf{r}_2\!\in\!\mathbb{C}^M$ at user 2 are given by
\begin{equation}
    \mathbf{r}_i=\mathbf{H}_i\,\mathbf{s}_{\Sigma}+\mathbf{n}_i,\quad{i}\in\{1,2\},
\end{equation}
where $\mathbf{n}_i\!\in\!\mathbb{C}^M$ represents the vector of complex additive white Gaussian noise (AWGN), additively distorting the received signal at user $i$, with each component having variance $\sigma^2$, also denoted as $N_0$. Further, in downlink NOMA transmission, the transmitted total signal $\mathbf{s}_{\Sigma}$ is subjected to the channel fading characterized by the diagonal channel matrix $\mathbf{H}_i\!=\!\sqrt{L_i}\diag({h}_{i,1},{h}_{i,2},\cdots,\allowbreak{h}_{i,M})\!\in\!\mathbb{C}^{M \times M}$. The path-loss (PL) coefficient is typically given by $L_i = (1 + d_i)^{-\xi}$, where $\xi$ is the path-loss exponent, commonly taken as $\xi=2$, and $d_i$ is the distance in meters between user $i$ and the base station. Additionally, under Rayleigh fading conditions, the channel coefficients $h_{i,m}$ for $1\!\leq\!m\!\leq\!M$ are characterized as independent and identically distributed (i.i.d.) complex Gaussian random variables, such that $h_{i,m}\!\sim\!\mathcal{CN}(0,1)$, where $\mathcal{CN}(0,1)$ denotes a circularly symmetric complex normal distribution with zero mean and unit variance. In contrast, for no-fading conditions, the channel coefficients simplify to $h_{i,m}\!=\!1$ for all $i\!\in\!\{1,2,\ldots,M\}$, representing a deterministic unity gain across all channels.

%%
%% New Paragraph
%%
In addition, for detection in NOMA setting, the strong user, defined as the one with the higher channel gain, employs successive interference cancellation (SIC) to decode its signal. The SIC process begins by decoding the signal of the weak user, characterized by the lower channel gain, and subsequently subtracting it to isolate the desired component. Let user 1 be designated as the strong user, such that their channel gain satisfies the inequality $\norm{\mathbf{H}_1}^2\!>\!\norm{\mathbf{H}_2}^2$. Under the assumption that the channel state information (CSI) is perfectly available, user 1 obtains its SIC-refined observation $\mathbf{r}_1^\mathrm{SIC}\!\in\!\mathbb{C}^M$, that is
\begin{equation}\label{eq:sic}
    \mathbf{r}_1^\mathrm{SIC}=\mathbf{r}_1-\sqrt{\alpha_2{P}}\,\mathbf{H}_1\widehat{\mathbf{s}}_{2\mid{1}},
\end{equation}
where $\widehat{\mathbf{s}}_{2\mid{1}}\!\in\!\mathbb{C}^M$ represents the reconstructed symbol block of user 2 as recovered by user 1. To recover $\widehat{\mathbf{s}}_{2\mid{1}}$, user 1 performs channel equalization and demodulation with hard-decision decoding on $\mathbf{r}_1$, yielding the bit sequence $\widehat{\mathbf{c}}_{2\mid{1}}\!\in\!\mathbb{F}_2^N$, which is then modulated to produce $\widehat{\mathbf{s}}_{2\mid{1}}$. After interference cancellation, user 1 applies channel equalization and demodulation to $\mathbf{r}_1^\mathrm{SIC}$ to retrieve its own bit sequence $\widehat{\mathbf{c}}_1\!\in\!\mathbb{F}_2^N$. In contrast, user 2, being the weak user, directly decodes its signal from $\mathbf{r}_2$ while treating the strong user’s signal as noise, resulting in the bit sequence $\widehat{\mathbf{c}}_2\!\in\!\mathbb{F}_2^N$. In practice, user 2 is typically allocated more transmission power to enhance the reliability of its signal decoding such that the approach not only ensures robust recovery for the weak user but also promotes fairness in resource allocation \cite{physical2023pakravan}.

%%   _________    ___.                         __  .__               
%%  /   _____/__ _\_ |__   ______ ____   _____/  |_|__| ____   ____  
%%  \_____  \|  |  \ __ \ /  ___// __ \_/ ___\   __\  |/  _ \ /    \ 
%%  /        \  |  / \_\ \\___ \\  ___/\  \___|  | |  (  <_> )   |  \
%% /_______  /____/|___  /____  >\___  >\___  >__| |__|\____/|___|  /
%%         \/          \/     \/     \/     \/                    \/ 
\subsection{Application of GRAND in NOMA system}\label{subsec:application_of_GRAND}
\IEEEcompact
Due to transmission distortions, the recovered bits $\widehat{\mathbf{c}}$ after demodulation may differ from the originally transmitted bits $\mathbf{c}$. This deviation can be modeled by an additive binary noise vector $\mathbf{e}\in\mathbb{F}_2^N$, such that $\widehat{\mathbf{c}}=\mathbf{c}\oplus\mathbf{e}$, whose inversion yields $\mathbf{e}=\widehat{\mathbf{c}}\oplus\mathbf{c}$. Upon searching for the most likely codeword $\mathbf{c}\in\mathcal{C}$ directly, a maximum-likelihood (ML) decoder can equivalently search for the most likely error pattern\cite[Eqs.~(4)~and~(5)]{bozkurt2024unlocking}
\begin{subequations}\label{Eq:MLDecoderSimplification}
\setlength\arraycolsep{1.4pt}
\begin{eqnarray}
\label{Eq:MLDecoderSimplificationA}
\mathbf{c}^{\mathrm{ML}}
    &\triangleq&\argmax_{\mathbf{c}\in \mathcal{C}}
        \Pr(\text{received}~\widehat{\mathbf{c}}\mid\text{transmitted}~\mathbf{c}),\\
\label{Eq:MLDecoderSimplificationE}
    &=&\argmax_{\mathbf{c}\in \mathcal{C}}
        \Pr(\mathbf{e}=\widehat{\mathbf{c}}\oplus\mathbf{c}),   
\end{eqnarray}
\end{subequations}
which actively seeks to estimate the optimal binary additive noise, maximizing the probability of successful CRC decoding, thus enabling the recovery of the transmitted bits with maximum accuracy. Therefore, it leads to the noise-centric decoding strategy known as GRAND \cite{duffy2019capacity,duffy2022ordered,bozkurt2024unlocking}, which systematically ranks all possible binary noise sequences from the most likely to the least likely, and sequentially subtracts them from the received noisy bit sequence until a valid codeword is identified, which is then output as the decoding result. Further, let $\mathcal{E}=\{\mathbf{e}_1, \mathbf{e}_2, \mathbf{e}_3,\dots\}\!\subseteq\!\mathbb{F}_2^N$ denote the set of candidate binary noise sequences to be tested such that $\Pr(\mathbf{e}_1)\ge\Pr(\mathbf{e}_2)\ge\Pr(\mathbf{e}_3)\ge\cdots\Pr(\mathbf{e}_{2^N})$. For a received bit sequence $\widehat{\mathbf{c}}\in\mathbb{F}_2^N$ after demodulation of $\mathbf{r}$ symbols, the decoding output of classical hard-decision GRAND is readily given by 
\begin{equation}\label{eq:c_grand}
\mathbf{c}^{\mathrm{GRAND}}=\widehat{\mathbf{c}}\oplus\mathbf{e}^\mathrm{GRAND},    
\end{equation}
which is the transmitted bits detected by GRAND decoder, where $\mathbf{e}^\mathrm{GRAND}$ denotes the detected error pattern (binary noise sequence) at which the GRAND decoding process terminates. Building on prior GRAND frameworks, the detection of $\mathbf{e}^\mathrm{GRAND}$ is formalized as \cite[Eq. (11)]{bozkurt2024unlocking}  
\begin{equation}\label{eq:e_grand}  
\mathbf{e}^\mathrm{GRAND}=
    \argmax_{\mathbf{e}\in\mathcal{E}}
        \Pr(\mathbf{e}\mid\CRC(\widehat{\mathbf{c}}\oplus\mathbf{e})~\text{succeeds}).  
\end{equation}

%%
%% New Paragraph
%%
The GRAND decoder, as described in equations \eqref{eq:c_grand} and \eqref{eq:e_grand}, can precisely correct bit errors (binary additive noise) in the estimated codewords $\widehat{\mathbf{c}}_1$ and $\widehat{\mathbf{c}}_2$, which are obtained through the demodulation of the received signals $\mathbf{r}_1^\mathrm{SIC}$ and $\mathbf{r}_2$ corresponding to user 1 and user 2, respectively. Conventionally, once GRAND is applied to these bit sequences, the corrected codewords $\mathbf{c}^\mathrm{GRAND}_1$ and $\mathbf{c}^\mathrm{GRAND}_2$ can be readily obtained. To improve decoding for user 1, we propose using GRAND within the SIC process. In this approach, user 1 applies GRAND to the bit sequence $\widehat{\mathbf{c}}_{2|1}$, which represents the estimated codeword of user 2 from the received symbols $\mathbf{r}_1$, obtaining the decoding output 
\begin{equation}
\!\!\mathbf{c}^{\mathrm{GRAND}}_{2\mid{1}}\!=\!
    \widehat{\mathbf{c}}_{2\mid{1}}\oplus
        \argmax_{\mathbf{e}\in\mathcal{E}}
            \Pr(\mathbf{e}\!\mid\!\CRC(\widehat{\mathbf{c}}_{2\mid{1}}\oplus\mathbf{e})~\text{succeeds}),\!\! 
\end{equation}
which yields an accurate reconstruction of the interfering codeword, which is then modulated to produce $\widehat{\mathbf{s}}_{2|1}$ and subtracted from $\mathbf{r}_1$ to enable effective interference cancellation. To the best of our knowledge, this is the first time GRAND has been used for both decoding and interference cancellation in a CRC-aided multi-user setting. In addition, we show that our framework is flexible—allowing a soft-decision version of GRAND to be smoothly integrated into the SIC process without any changes to the algorithm. This integration lets the decoder make use of soft channel information while still keeping the computation efficient.

%%%%%%%%%%%%%%%%%%%%%%%%%%%%%%%%%%%%%%%%%%%%%%%%%%%%%%%%%%%%%%%%%%%%%%%%%%%%%%%%%%%%%%%%%%%%%%%%%%%%%%%%%%%%%%%%%%%%%%%%%%%%
%%   _________              __  .__               
%%  /   _____/ ____   _____/  |_|__| ____   ____  
%%  \_____  \_/ __ \_/ ___\   __\  |/  _ \ /    \ 
%%  /        \  ___/\  \___|  | |  (  <_> )   |  \
%% /_______  /\___  >\___  >__| |__|\____/|___|  /
%%         \/     \/     \/                    \/ 
\section{Performance Analysis}\label{Sec:PerformanceAnalysis}
\IEEEcompact
To examine the scenario where GRAND is used prior to applying SIC, we start by analyzing the received signal $\mathbf{r}_{2|1}$ at user 1. Demodulation on $\mathbf{r}_{2|1}$ yields the estimated codeword $\widehat{\mathbf{c}}_{2|1}$. If GRAND is applied to this estimate and successfully correct all bit errors, the corrected codeword becomes $\widehat{\mathbf{c}}_{2\mid{1}}\oplus\mathbf{e}^{\mathrm{GRAND}}_{2\mid{1}}
\!\implies\!\mathbf{c}^{\mathrm{GRAND}}_{2\mid{1}}\allowbreak\!\implies\!\mathbf{c}_2$, 
where $\mathbf{e}^{\mathrm{GRAND}}_{2|1} $ is the binary noise sequence identified by GRAND decoder. In accordance with \eqref{eq:sic}, the contribution of user 2 is perfectly canceled out from $ \mathbf{r}_{1}$. As a result, the interference-free signal becomes
\begin{equation}
    \mathbf{r}_1^\mathrm{SIC}=\sqrt{\alpha_1{P}}\,\mathbf{H}_1\mathbf{s}_1 + \mathbf{n}_1.
\end{equation}
which shows that, assuming perfect CSI is available, the signal $\mathbf{s}_2$ of user 2 is entirely eliminated from the received signal $\mathbf{r}_1$, leaving only the signal $\mathbf{s}_1$ and additive noise $\mathbf{n}_1$. On the other hand, the crucial limitation of GRAND decoder is its vulnerability to undetectable errors. Despite being designed to detect errors, GRAND decoder may fail to identify certain types of errors. For example, if the  error pattern (additive binary noise) is an element of the CRC codebook $\mathcal{C}$, it cannot be detected by GRAND decoder, yielding $\widehat{\mathbf{c}}_{2\mid{1}}\oplus\mathbf{e}^{\mathrm{GRAND}}_{2\mid{1}}
\!\implies\!\allowbreak\mathbf{c}^{\mathrm{GRAND}}_{2\mid{1}}\!\implies\!\allowbreak\mathbf{c}_2\oplus\widehat{\mathbf{e}}^{\mathrm{GRAND}}_{2\mid{1}}$, where $\widehat{\mathbf{e}}^{\mathrm{GRAND}}_{2\mid{1}}$ is the undetectable error pattern passing CRC check, meaning it is not flagged as errors. Hence, $\mathbf{c}^{\mathrm{GRAND}}_{2\mid{1}}$ is modulated as $\mathbf{s}^{\mathrm{GRAND}}_{2\mid{1}}$ and then canceled out from $\mathbf{r}_{1}$ as
\begin{equation}
    \mathbf{r}_1^\mathrm{SIC}= 
        \sqrt{P\alpha_1}\,\mathbf{H}_1\mathbf{s}_1+
            \sqrt{P\alpha_2}\,\mathbf{H}_1
                (\mathbf{s}_2-\mathbf{s}^{\mathrm{GRAND}}_{2\mid{1}})+\mathbf{n}_1.
\end{equation}
Due to space constraints, the derivation steps for the bit error probability are omitted. Accordingly, the bit error probability for user 1 under BPSK modulation is given by
\begin{multline}
        P^{\text{user 1}}_{\mathrm{BER}}=P_{\mathrm{UE}}\,\mathbb{E}\Biggl[Q\Biggl(
    2\sqrt{
    \frac{2\alpha_1{P}\bigl\lVert\mathbf{H}_1 \mathbf{s}_1\bigr\rVert^2 }
    {4\alpha_2{P}\bigl\lVert\mathbf{H}_1
           \mathbf{s}_2\bigr\rVert^2 + N\sigma^2 }}\Biggr)\Biggr]+\\
          (1-P_{\mathrm{UE}})\mathbb{E}\Biggl[Q\Biggl(2\sqrt{\frac{\alpha_1{P}\bigl\lVert\mathbf{H}_1 \mathbf{s}_1\bigr\rVert^2 }
    {N\sigma^2 }}\Biggr)\Biggr].
\end{multline}
where $Q(x)$ denotes Gaussian's Q function \cite[Eq. (2)]{new2003chiani}, and $P_\mathrm{UE}=\mathbb{E}\bigl[\widehat{\mathbf{e}}^{\mathrm{GRAND}}_{2\mid{1}}\bigl]$ denotes the average of undetectable error. Similarly, the bit error probability for user 2 using BPSK modulation is obtained as 
\begin{equation}
    P^{\text{user 2}}_{\mathrm{BER}}=
        \mathbb{E}\Biggl[Q\Biggl(2\sqrt{\frac{\alpha_2{P}\bigl\lVert\mathbf{H}_2 \mathbf{s}_2\bigr\rVert^2 }
    {N\sigma^2}}\Biggr)\Biggr].
\end{equation}
If the GRAND decoder can correct up to $L$ bit errors, an upper bound on the block error rate (BLER) for user $i\in\{1,2\}$, representing the probability of uncorrectable errors (i.e., when more than $L$ errors occur), can be calculated as  
\begin{equation}
P^{\text{user}~i}_{\mathrm{BLER}}\leq 1-\sum_{\ell=0}^{L}
    \binom{N}{\ell}
        \bigl(P^{\text{user}~i}_{\mathrm{BER}}\bigr)^\ell\bigl(1-P^{\text{user}~i}_{\mathrm{BER}}\bigr)^{N-\ell},
\end{equation}
where $\binom{N}{l}$ denotes the binomial coefficient. 

%%%%%%%%%%%%%%%%%%%%%%%%%%%%%%%%%%%%%%%%%%%%%%%%%%%%%%%%%%%%%%%%%%%%%%%%%%%%%%%%%%%%%%%%%%%%%%%%%%%%%%%%%%%%%%%%%%%%%%%%%%%%
%%   _________              __  .__               
%%  /   _____/ ____   _____/  |_|__| ____   ____  
%%  \_____  \_/ __ \_/ ___\   __\  |/  _ \ /    \ 
%%  /        \  ___/\  \___|  | |  (  <_> )   |  \
%% /_______  /\___  >\___  >__| |__|\____/|___|  /
%%         \/     \/     \/                    \/ 
\section{Performance Results}\label{Sec:PerformanceAnalysisAndSimulationResults}
In this section, we present only the BER performance results of the proposed communication system as a function of transmitted bit energy over noise variance (i.e., $E_b/N_0$). Additional block error rate (BLER) results were also obtained and exhibit trends parallel to the BER outcomes; however, they have not been presented in this paper due to page limitations. We evaluate all the previously described GRAND-NOMA scenarios over AWGN and Rayleigh fading channels. During the simulations, information bit blocks of length $K = 116$ are encoded to $N = 128$ bit length codewords via CRC-12 with Koopman polynomial $0x8f3$. Binary phase-shift keying is the chosen modulation scheme. We define user 1 as the user that is closer to the base station, so that $d_1 \leq d_2$. For the transmission power, we fix $P = 1$. The power allocation is set as \( \alpha_2 = 0.75 \) for the far user (user 2) and \( \alpha_1 = 0.25 \) for the near user (user 1). We assume that both users have perfect CSI of their respective channels for channel equalization and a perfect estimate of the noise variance for obtaining the log likelihood ratio (LLR) of their received signals. We use the classical hard decision GRAND and soft decision 1-line ORBGRAND, referred to as ORBGRAND in the remainder of this section, as decoders \cite{galligan2023block}. 

The system is simulated under three different scenarios: Pure NOMA, GRAND-NOMA, and GRAND-NOMA with assistance. In the Pure NOMA scenario, the information bits of user 2 are obtained directly through channel equalization and demodulation. User 1 on the other hand applies SIC to its received signal and again obtains the information bits via channel equalization and demodulation. The GRAND-NOMA scenario refers to the case where both user 1 and 2 employ GRAND decoding to recover their respective information bits after demodulating their received signals. In the GRAND-NOMA with assistance scenario, in addition to the previous scenarios, user 1 performs the SIC process while also utilizing GRAND decoding to decode the other user's information bits (as explained in Sec. \ref{subsec:application_of_GRAND}). The operation of user 2 is the same as in GRAND-NOMA. 

\begin{figure}[t]
  \centering
  \begin{minipage}{0.4\textwidth}
    \includegraphics[width=\textwidth,trim=0.0cm 0.0cm 0.0cm 0.0cm, clip, keepaspectratio]{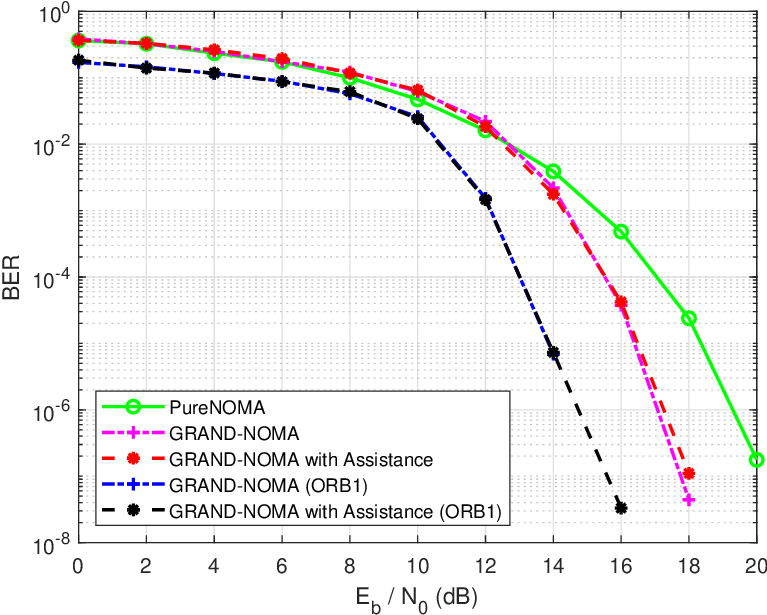}
    \subcaption{Over AWGN channel (no-fading).}
    \label{Fig:user1_BER_AWGN}
  \end{minipage}
  \\[2mm]%\hfill
  \begin{minipage}{0.4\textwidth}
    \includegraphics[width=\textwidth,trim=0.0cm 0.0cm 0.0cm 0.0cm, clip, keepaspectratio]{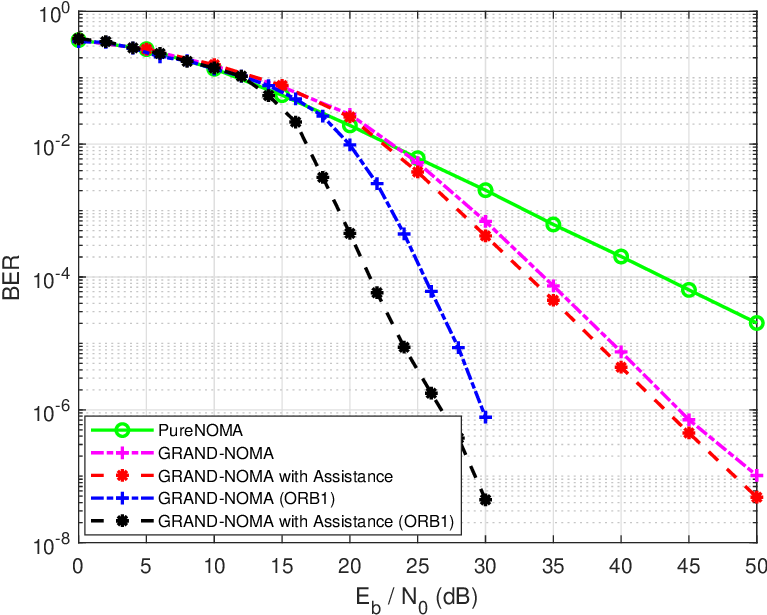}
    \subcaption{Over Rayleigh fading channel.}
    \label{Fig:user1_BER_Rayleigh}
  \end{minipage}
  \\[1mm]
  \setlength{\belowcaptionskip}{-5mm}
  \caption{\small BER performance of user 1 for different $E_b/N_0$ under Pure NOMA, GRAND-NOMA, and GRAND-NOMA with assistance using GRAND and ORBGRAND.}
  \label{Fig:user1_BER}
  %\vspace{-5mm}
\end{figure}

\begin{figure}[h]
  \centering
  \begin{minipage}{0.4\textwidth}
    \includegraphics[width=\textwidth,trim=0.0cm 0.0cm 0.0cm 0.0cm, clip, keepaspectratio]{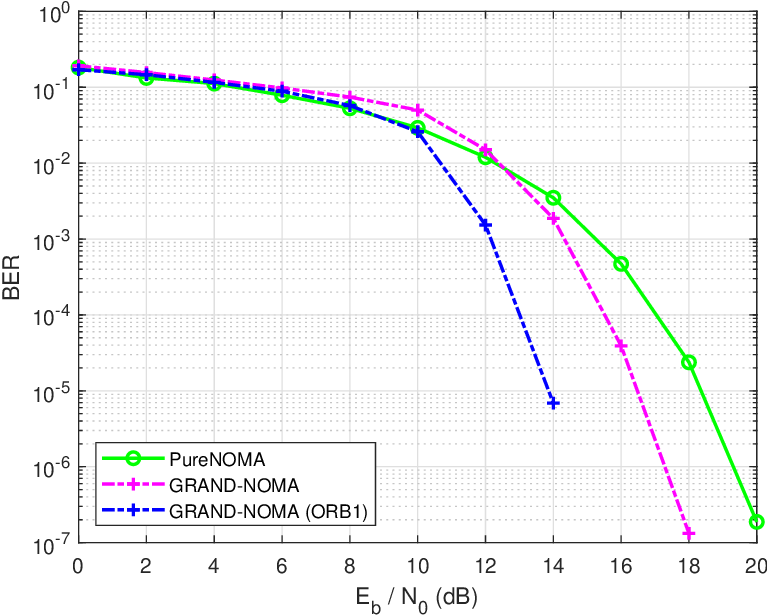}
    \subcaption{Over AWGN channel (no-fading).}
    \label{Fig:user2_BER_AWGN}
  \end{minipage}
  \\[2mm]%\hfill
  \begin{minipage}{0.4\textwidth}
    \includegraphics[width=\textwidth,trim=0.0cm 0.0cm 0.0cm 0.0cm, clip, keepaspectratio]{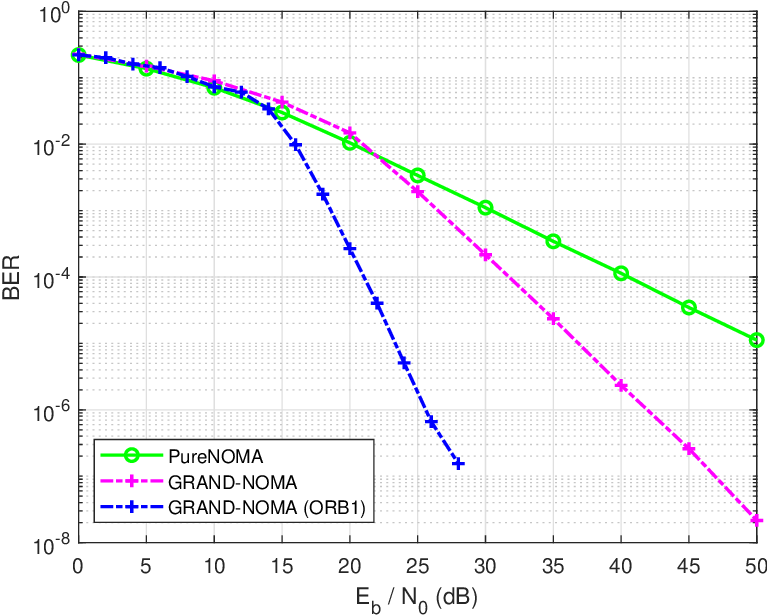}
    \subcaption{Over Rayleigh fading channel.}
    \label{Fig:user2_BER_Rayleigh}
  \end{minipage}
  \\[1mm]
  \setlength{\belowcaptionskip}{-5mm}
  \caption{\small BER performance of user 2 for different $E_b/N_0$ under Pure NOMA and GRAND-NOMA using GRAND and ORBGRAND.}
  \label{Fig:user2_BER}
  %\vspace{2mm}
\end{figure}

\begin{figure}[h]
  \centering
  \begin{minipage}{0.4\textwidth}
    \includegraphics[width=\textwidth,trim=0.0cm 0.0cm 0.0cm 0.315cm, clip, keepaspectratio]{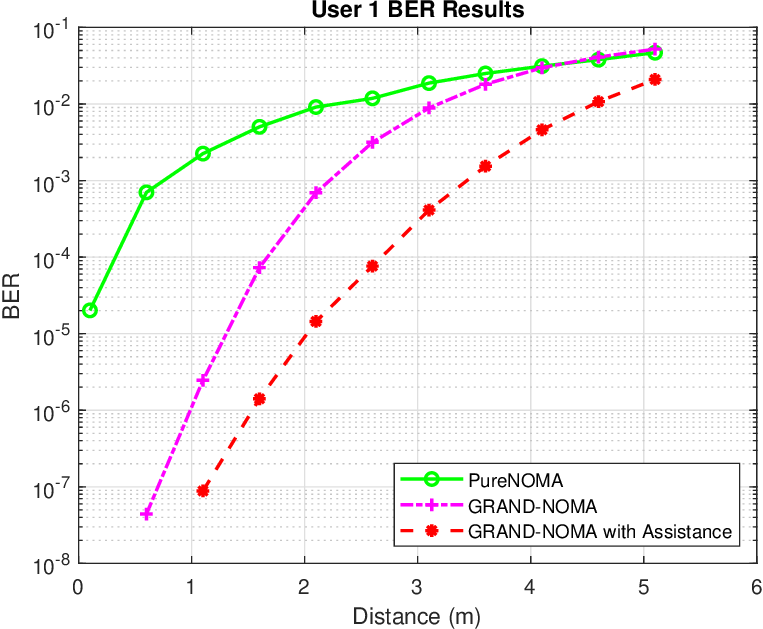}
  \end{minipage}
  \setlength{\belowcaptionskip}{-5mm}
  \caption{\small BER performance of user 1 versus distance under Pure NOMA, GRAND-NOMA, and GRAND-NOMA with Assistance.}
  \label{Fig:distance_User1}
  %\vspace{-5mm}
\end{figure}

In this section, we provide an in-depth discussion of the simulation results. Three distinct scenarios involving both user 1 and user 2 were simulated using the GRAND and ORBGRAND algorithms. In the simulations, user 1 and 2 are placed at a distance of 1 meter since \(E_b/N_0\) values are calculated in the transmitter.  The simulation results for user 1 are presented in Fig. \ref{Fig:user1_BER}, where Figure \ref{Fig:user1_BER_AWGN} illustrates the outcomes for the AWGN channel and Figure \ref{Fig:user1_BER_Rayleigh} for the Rayleigh channel. In the AWGN channel, decoding user 1’s signal using GRAND-NOMA with ORBGRAND yields the best performance across all \(E_b/N_0\) values. At an $E_b/N_0$ of 16 dB, GRAND NOMA with Assistance employing ORBGRAND achieves a BER of $2 \times 10^{-8}$, whereas GRAND NOMA and Pure NOMA yield BER values of $4 \times 10^{-5}$ and $4 \times 10^{-4}$, respectively. In order for GRAND-NOMA and Pure NOMA to provide a BER close to $2 \times 10^{-8}$, they require an SNR increment of more than 2 and 4 dB, respectively. However, GRAND shows a performance-enhancing effect to Pure NOMA beyond a certain dB threshold, and it perform worse than ORBGRAND. In the AWGN channel, the GRAND-NOMA Assistance method does not offer a significant advantage over the GRAND-NOMA approach. However, in the Rayleigh channel, when the GRAND-NOMA with Assistance method is implemented with ORBGRAND, it provides a substantial improvement in error performance. In contrast, using the assistance method with GRAND does not result in a major enhancement. For instance in Rayleigh channel, at an \(E_b/N_0\) of 30 dB, Pure NOMA attains a BER of $2 \times 10 ^{-3}$, the methods incorporating GRAND reach approximately $4 \times 10^{-4}$, GRAND-NOMA with ORBGRAND achieves a BER of $8 \times 10^{-7}$, and GRAND-NOMA with Assistance using ORBGRAND yields a BER of $4 \times 10^{-8}$. These results indicate that, for user 1, the ORBGRAND-based techniques deliver the best performance in the AWGN channel, while in the Rayleigh channel the GRAND-NOMA with Assistance method using ORBGRAND produces the optimal outcomes. This is an expected result, since ORBGRAND takes bit reliabilities into account when decoding the information from the other user.

% Fig. \ref{Fig:user2_BER} presents the results obtained for user 2. Simulation studies were conducted under both AWGN and Rayleigh channel conditions. Since the far user does not decode the other user's information bits, the GRAND-NOMA with Assistance method is omitted from the simulation for user 2. The results demonstrate that, for user 2, the best performance in both AWGN and Rayleigh channels is achieved when using ORBGRAND—with a more pronounced gain observed in the Rayleigh channel. Additionally, GRAND outperforms Pure NOMA after surpassing a certain dB threshold in both channels. Overall, the findings confirm that ORBGRAND delivers the optimal performance for user 2 in both channel conditions. This outcome is expected, as the ORBGRAND method ranks error sequences according to soft information inputs and accounts for the amount of channel distortion. Figure~\ref{Fig:user2_BER_AWGN} illustrates the bit error rate (BER) performance for User 2 in an AWGN channel. At an $E_b/N_0$ value of 14 dB, GRAND-NOMA with ORBGRAND achieves a BER of $7 \times 10^{-6}$, GRAND-NOMA with classical GRAND results in a BER of $2 \times 10^{-3}$, while PureNOMA yields a BER of $4 \times 10^{-3}$. To achieve a bit error rate (BER) around $7 \times 10^{-6}$, GRAND-NOMA and Pure NOMA require an SNR enhancement exceeding 2 dB and 4 dB, respectively.

\figref{Fig:user2_BER} shows the results for user 2 under both AWGN and Rayleigh channels. Since the far user does not decode the other user's data, GRAND-NOMA with Assistance is excluded from simulations for user 2. The results reveal that ORBGRAND achieves the best performance for user 2 across both channels, with a more substantial gain in Rayleigh fading. Moreover, GRAND surpasses Pure NOMA after a certain SNR threshold in both scenarios. These outcomes align with ORBGRAND’s design, which ranks error sequences using soft information and accounts for channel distortion. \figref{Fig:user2_BER_AWGN} presents BER performance for user 2 in AWGN. At $ E_b/N_0 = 14$ dB, ORBGRAND reaches a BER of $7\times{10}^{-6}$, while classical GRAND gives $2 \times 10^{-3}$, and Pure NOMA yields $4\times{10}^{-3}$. To attain $ 7 \times 10^{-6}$ BER, GRAND and Pure NOMA require over $2$ dB and $4$ dB higher SNR, respectively.

\begin{figure}[t!]
  \centering
  \begin{minipage}{0.4\textwidth}
    \includegraphics[width=\textwidth,trim=0.0cm 0.0cm 0.0cm 0.332cm, clip, keepaspectratio]{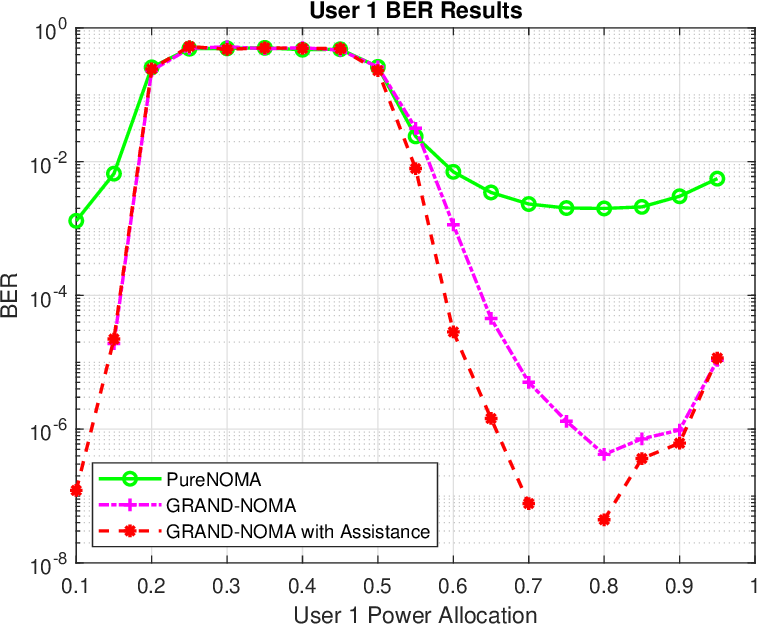}
    \subcaption{BER of user 1 with respect to power allocation $\alpha_1$.}
    \label{Fig:power_User1_BER}
  \end{minipage}
  \\[2mm]%\hfill
  \begin{minipage}{0.4\textwidth}
    \includegraphics[width=\textwidth,trim=0.0cm 0.0cm 0.0cm 0.332cm, clip, keepaspectratio]{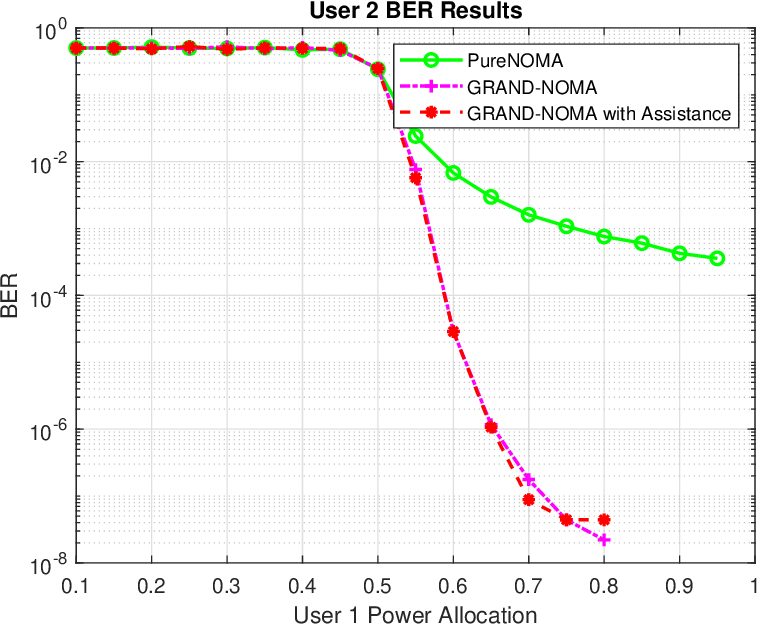}
    \subcaption{BER of user 2 with respect to power allocation $\alpha_2$.}
    \label{Fig:power_User2_BER}
  \end{minipage}
  \\[1mm]
  \setlength{\belowcaptionskip}{-5mm}
  \caption{\small BER comparison of Pure NOMA, GRAND-NOMA, and GRAND-NOMA with assistance with respect to power allocations.}
  \label{Fig:power_User1_User2}
  %\vspace{-5mm}
\end{figure}

\figref{Fig:power_User1_User2} displays the BER curves of both users to their power allocation constants. For the results of this figure, the channel has Rayleigh fading, we set $d_1 = d_2 = 1$, $E_b/N_0 = 30$ dB and use ORBGRAND. Looking at the BER results for user 1 in \figref{Fig:power_User1_BER}, it is seen that the curve of GRAND-NOMA with assistance displays the lowest BER values at each $\alpha_1$ value while Pure NOMA displays the highest BER values. As a common point, we observe that all the curves pertain to the same trend of increasing and decreasing at certain ranges. We see that for $0.1 \leq \alpha_1 \leq 0.5$, the BER is in the vicinity of $0.5$, except for when $\alpha_1$ is close to $0.1$. The high BER can be explained through the fact that when $0.1 \leq \alpha_1 \leq 0.5$ is not sufficiently close to $0.1$, the signal of user 1 causes critical interference with that of user 2, leading to inaccurate SIC and thus poor performance. Obviously, when $\alpha_1$ is close to $0.1$, there is negligible interference, hence SIC gives the signal of user 1 precisely, resulting in lower BER. For the other half of the plot, i.e., $0.5 < \alpha_1 \leq 0.95$, all curves exhibit a concave shape where the minimum BER value is attained around $\alpha_1 = 0.75$. In the range $0.5 < \alpha_1 \leq 0.7$, as $\alpha_1$ increases, the interference caused by unsuccessful SIC is suppressed since this interfering term is multiplied with $\alpha_2 = 1- \alpha_1$ according to \eqref{eq:sic}. This leads to the BER performance increasing. However, as $\alpha_1$ exceeds $0.75$ and nears $0.95$, a degradation in the BER performance is observed. This is due to the SIC process now starting to cancel an interference term that is almost purely constructive, which overwhelms the aforementioned effect of increasing $\alpha_1$ that improves BER. Put simply, the term $\widehat{\mathbf{s}}_{2\mid{1}}$ starts to become so similar to $\mathbf{s}_1$ such that despite the increasing $\alpha_1$, the signal resulting from SIC is more corrupted, causing the BER to suffer. The curves in Fig. \ref{Fig:power_User2_BER} show the BER results of user 2 for differing $\alpha_2$. Because user 2 does not perform SIC, the curves' shapes are much simpler compared to those of Fig. \ref{Fig:power_User1_BER}. Observing Fig. \ref{Fig:power_User2_BER}, we see that with increasing $\alpha_2$ BER decreases. For $0.1 \leq \alpha_2 \leq 0.5$, the signal of user 1 completely overshadows the signal of user 2 since $\alpha_1 \geq \alpha_2$, causing the BER to be approximately $0.5$. In the range $0.5 < \alpha_2 \leq 0.95$ however, we have $\alpha_2 > \alpha_1$, and naturally, the BER decreases as $\alpha_2$ increases. 

In addition, simulations were performed by varying the distance of the user 1, $d_1$. Fig. \ref{Fig:distance_User1} presents the BER result for the user 1 as a function of $d_1$. In these simulations, the GRAND algorithm was employed, while the distance of user 2 was fixed at $3$ meters. Although the user 1 transitions into the user 2 region beyond 3 meters, the decoding and SIC process remains unchanged. The system was tested at an SNR of 30 dB $(E_b/N_0)$. According to the results, as user 1 moves closer to the transmitter, the performance gap between the proposed GRAND-NOMA with assistance and Pure NOMA increases. However, as the distance increases up to 5 meters, the performance differences among the three algorithms diminish. Nonetheless, across all distances, GRAND-NOMA with assistance consistently outperforms the other methods. Specifically, at a distance of $1.6$ meters, GRAND-NOMA with assistance achieved a BER of $1.41 \times 10^{-6}$, while GRAND-NOMA and Pure NOMA yielded BERs of $7.3 \times 10^{-5}$ and $5.03 \times 10^{-3}$, respectively.

%%%%%%%%%%%%%%%%%%%%%%%%%%%%%%%%%%%%%%%%%%%%%%%%%%%%%%%%%%%%%%%%%%%%%%%%%%%%%%%%%%%%%%%%%%%%%%%%%%%%%%%%%%%%%%%%%%%%%%%%%%%%
%%   _________              __  .__               
%%  /   _____/ ____   _____/  |_|__| ____   ____  
%%  \_____  \_/ __ \_/ ___\   __\  |/  _ \ /    \ 
%%  /        \  ___/\  \___|  | |  (  <_> )   |  \
%% /_______  /\___  >\___  >__| |__|\____/|___|  /
%%         \/     \/     \/                    \/ 
\section{Conclusion}\label{Sec:Conclusion}
This paper demonstrates a novel two-user downlink power-domain NOMA framework that seamlessly combines CRC-aided GRAND with SIC, introducing a noise-centric and FEC-free decoding strategy. By leveraging GRAND’s efficient error-pattern search and integrating CRC into the decoding process, the proposed scheme effectively mitigates error propagation. Simulation results over AWGN and Rayleigh fading channels show that our approach significantly enhances the error performance compared to conventional NOMA. These advancements position the CRC-aided GRAND-NOMA framework as a strong candidate for Ultra Reliable Low Latency Communications (URLLC) and massive machine-type communications (mMTC) applications in next-generation 5G and 6G wireless networks, where spectral efficiency and robust multi-user support are critical.

% In this paper, we introduce a novel two-user downlink power-domain NOMA scheme that integrates CRC‑aided GRAND with SIC. By exploiting GRAND’s noise-centric error‑pattern ranking and CRC checks, our design eliminates the need for separate FEC codes and systematically overcomes the error‑propagation issues inherent to conventional SIC. Extensive simulations over AWGN and Rayleigh fading channels demonstrate that the CRC‑aided GRAND‑NOMA framework consistently outperforms the existing traditional NOMA transmission.

%\section*{Acknowledgment}

%\bibliography{IEEEfull,zor_bozkurt_yilmaz_grand_crc_noma}
\bibliography{IEEEabrv,zor_bozkurt_yilmaz_grand_crc_noma}

\begin{thebibliography}{10}
\providecommand{\url}[1]{#1}
\csname url@samestyle\endcsname
\providecommand{\newblock}{\relax}
\providecommand{\bibinfo}[2]{#2}
\providecommand{\BIBentrySTDinterwordspacing}{\spaceskip=0pt\relax}
\providecommand{\BIBentryALTinterwordstretchfactor}{4}
\providecommand{\BIBentryALTinterwordspacing}{\spaceskip=\fontdimen2\font plus
\BIBentryALTinterwordstretchfactor\fontdimen3\font minus \fontdimen4\font\relax}
\providecommand{\BIBforeignlanguage}[2]{{%
\expandafter\ifx\csname l@#1\endcsname\relax
\typeout{** WARNING: IEEEtran.bst: No hyphenation pattern has been}%
\typeout{** loaded for the language `#1'. Using the pattern for}%
\typeout{** the default language instead.}%
\else
\language=\csname l@#1\endcsname
\fi
#2}}
\providecommand{\BIBdecl}{\relax}
\BIBdecl

\bibitem{survey2020makki}
B.~Makki, K.~Chitti, A.~Behravan, and M.-S. Alouini, ``{A Survey of NOMA: Current Status and Open Research Challenges},'' \emph{IEEE Open Journal of the Communications Society}, vol.~1, pp. 179--189, 2020.

\bibitem{duffy2019capacity}
K.~R. Duffy, J.~Li, and M.~M{\'e}dard, ``{C}apacity-{A}chieving {G}uessing {R}andom {A}dditive {N}oise {D}ecoding,'' \emph{IEEE Transactions on Information Theory}, vol.~65, no.~7, pp. 4023--4040, 2019.

\bibitem{duffy2022ordered}
K.~R. Duffy, W.~An, and M.~M{\'e}dard, ``Ordered {R}eliability {B}its {G}uessing {R}andom {A}dditive {N}oise {D}ecoding,'' \emph{IEEE Transactions on Signal Processing}, vol.~70, pp. 4528--4542, 2022.

\bibitem{multiuser2023yang}
K.~Yang, M.~M{\'e}dard, and K.~R. Duffy, ``{Multiuser Detection Using GRAND-Aided Macrosymbols},'' in \emph{ICC 2023-IEEE International Conference on Communications}.\hskip 1em plus 0.5em minus 0.4em\relax IEEE, 2023, pp. 4646--4651.

\bibitem{yang2023interference_aware}
------, ``Separating {I}nterferers from {M}ultiple {U}sers in {I}nterference {A}ware {G}uessing {R}andom {A}dditive {N}oise {D}ecoding {A}ided {M}acrosymbol,'' in \emph{MILCOM 2023-2023 IEEE Military Communications Conference (MILCOM)}.\hskip 1em plus 0.5em minus 0.4em\relax IEEE, 2023, pp. 643--648.

\bibitem{yang2024noma_grand_am}
------, ``{GRAND-AM} in uplink {MIMO NOMA} systems,'' in \emph{59th Annual Conf. on Info. Sciences and Systems (CISS)}.\hskip 1em plus 0.5em minus 0.4em\relax IEEE, 2025, pp. 1--6.

\bibitem{umar2024joint_sgrand_polar}
R.~Umar, A.~ul~Quddus, and Y.~Ma, ``Joint {CRC} {Aided Soft-GRAND for Decoding of Polar Codes over an Interference Channel with NOMA Protocol},'' in \emph{IEEE Wireless Communications and Networking Conference (WCNC)}.\hskip 1em plus 0.5em minus 0.4em\relax IEEE, 2024, pp. 1--6.

\bibitem{physical2023pakravan}
S.~Pakravan, J.-Y. Chouinard, X.~Li, M.~Zeng, W.~Hao, Q.-V. Pham, and O.~A. Dobre, ``{Physical Layer Security for NOMA Systems: Requirements, Issues, and Recommendations},'' \emph{IEEE Internet of Things Journal}, vol.~10, no.~24, pp. 21\,721--21\,737, 2023.

\bibitem{bozkurt2024unlocking}
B.~Bozkurt, E.~Zor, and F.~Yilmaz, ``{Unlocking Potential: Integrating Multihop, CRC, and GRAND for Wireless 5G-Beyond/6G Networks},'' in \emph{2024 6th International Conference on Communications, Signal Processing, and their Applications (ICCSPA)}.\hskip 1em plus 0.5em minus 0.4em\relax IEEE, 2024, pp. 1--6.

\bibitem{new2003chiani}
M.~Chiani, D.~Dardari, and M.~K. Simon, ``New exponential bounds and approximations for the computation of error probability in fading channels,'' \emph{IEEE Transactions on Wireless Communications}, vol.~2, no.~4, pp. 840--845, 2003.

\bibitem{galligan2023block}
K.~Galligan, M.~M{\'e}dard, and K.~R. Duffy, ``Block turbo decoding with {ORBGRAND},'' in \emph{57th Annual Conf. on Inf. Sci. Systems (CISS)}, 2023.

\end{thebibliography}
\bibliographystyle{IEEEtran}

\end{document}